\documentclass[10pt,showpacs,preprintnumbers,amsmath,eqsecnum,
amssymb,twocolumn,aps,prd,nofootinbib,
a4paper,superscriptaddress]{revtex4}
\pdfoutput=1

\usepackage{graphicx}

\def\be{\begin{equation}}
\def\ee{\end{equation}}

\def\bea{\begin{eqnarray}}
\def\eea{\end{eqnarray}}

\def\vp{{\varphi}}

\def\cs2{c_{\rm{s}}^2}

\def\U0{{\bar U_0}}

\def\bi{\begin{itemize}}
\def\ei{\end{itemize}}

\usepackage{epsfig}
\usepackage{graphicx,epsf}
\usepackage{color}
\usepackage{bm}
\usepackage{psfrag}

\def\be{\begin{equation}}
\def\ee{\end{equation}}
\def\beb{\begin{equation*}}
\def\eeb{\end{equation*}}
\def\bea{\begin{eqnarray}}
\def\eea{\end{eqnarray}}
\def\beab{\begin{eqnarray*}}
\def\eeab{\end{eqnarray*}}
\def\nn{\nonumber}

\def\vp{{{\varphi}}}

\def\cs2{c_{\rm{s}}^2}

\def \beg {\begin{enumerate}}
\def \en {\end{enumerate}}

\def\cs{c_{\rm{s}}^2}

\newcommand{\Gam}{\ensuremath{\Gamma^A_{\mathrm{m}}}}
\newcommand{\Gag}{\ensuremath{\Gamma^A_{\gamma}}}

\begin{document}
\title{Isocurvature Perturbations and Reheating in Multi-Field Inflation}
\author{Ian~Huston}
\email[]{i.huston@qmul.ac.uk}
\affiliation{Astronomy Unit, School of Physics and Astronomy, Queen Mary University of London,
Mile End Road, London, E1 4NS, UK}

\author{Adam J.~Christopherson}
\email[]{Adam.Christopherson@nottingham.ac.uk}
\affiliation{School of Physics and Astronomy, University of Nottingham, University Park,
Nottingham, NG7 2RD, UK}

\date{\today}

\begin{abstract} 
Inflationary models involving more than one scalar field naturally produce
isocurvature perturbations. However, while these are fairly well studied, less
is known about their evolution through the reheating
epoch, when the inflationary fields decay into the standard constituents of the 
present universe. In this paper, by modelling
reheating perturbatively, we calculate
the power spectrum of the non-adiabatic pressure perturbation in three
different inflationary models. We show that the isocurvature can grow large
initially, but decays faster than the pressure perturbations. When reheating ends,
the isocurvature is negligible for the double quadratic and 
double quartic inflationary models. For the product exponential potential,
which features large isocurvature at the end of inflation, the isocurvature
decays during reheating and is around five orders of magnitudes
smaller than the pressure perturbation at the end of reheating.
\end{abstract}


\maketitle

\section{Introduction}
\label{sec:intro}

The inflationary paradigm is incredibly successful in solving the original problems of big bang cosmology \cite{LLBook},
to such an extent that it is now firmly a part of standard cosmology.  However, despite this success,
we have no single, specific theory of inflation which fits the data better than all others.
 There has been a plethora of inflationary
models considered in the literature to date, ranging from the original and simplest single field, slow-roll potential
\cite{Guth:1980zm, Starobinsky:1980te, Linde:1981mu}, 
through more general models involving multiple scalar fields (see, e.g., \cite{Bassett:2005xm} and
references within), to more complicated scenarios involving scalar
fields with non-standard kinetic terms, such as k-inflation \cite{ArmendarizPicon:1999rj}.
It is anticipated that data obtained from current and future
experiments, such as observations of the cosmic microwave background (CMB) from the Planck satellite\cite{Planck},
will enable us to constrain the inflationary paradigm.

One signature which can be used to distinguish between different inflationary models
is the isocurvature, generated by the relative entropy perturbation between scalar fields.
This naturally occurs in any general system containing more than one field. Constraints on primordial isocurvature
come from observations of the CMB. However, to date, the data only constrain the ratio of isocurvature to adiabatic
perturbations to be of the order of $0.1$ \cite{WMAP7}. An interesting possibility for determining a favoured inflationary
model from its primordial isocurvature signature
is through the vorticity induced by non-adiabatic pressure at second order in perturbation theory 
\cite{vorticity, Christopherson:2010dw, Christopherson:2010ek}. The possibility of 
constraining isocurvature in this manner is discussed in more detail in Ref.~\cite{Huston:2011fr}.

In addition to the unknowns around the specific model of inflation, we do not know how the scalar
fields driving inflation decay in the standard model particles that exist in the universe today.
This process is dubbed `reheating' and there is currently no agreed upon mechanism
by which the universe reheats.

For models consisting of multiple scalar fields (excluding those where the isocurvature has decayed by the end
of inflation since these can then be treated as single field systems)
this process of reheating is of utmost importance. This is because 
isocurvature perturbations naturally `spoil' the standard picture of the early universe since the curvature
perturbation is no longer conserved on super-horizon scales. This makes it interesting, and important, to address how the
isocurvature remaining at the end of the inflationary phase evolves through a period of reheating. 

There are different methods of modelling the reheating phase, either by focussing on the particle interactions
and particle production during reheating (see, e.g., Ref.~\cite{Allahverdi:2010xz} and references therein),
or by considering, in a more classical manner,  
how the inflationary fields decay into the matter and radiation components of the standard cosmological model. 
It is the latter technique which we adopt in this article, using full cosmological perturbation theory 
\cite{Bardeen:1980kt, ks, mfb, MW2008, Malik:2008yp}.
Of course, if we consider the scalar fields decaying into
a single fluid, as has been done in some other work, the system will not be able to support any 
evolution of isocurvature.

Therefore, in this article we consider the decay of the inflationary scalar fields
into more than one species of fluid and
 investigate how reheating affects isocurvature perturbations generated during multi-field inflation.
Using results from Ref.~\cite{Huston:2011fr}, we model the reheating process by a system containing two scalar 
fields along with a matter and radiation fluids with decay constants converting the scalar fields into fluids. 
By modelling the system like this, we explore the scenario in a more general manner than in previous work, 
and consider the non-adiabatic pressure perturbations
between the radiation and the matter fluids after reheating is complete. In doing so, we can get a handle on whether it is 
common for isocurvature to survive through reheating. 

This paper is organised as follows: in the next section, we review the details of perturbative
reheating, and present the set of equations to be solved. In Section~\ref{sec:num} we review the 
numerical procedure. We present the results for the three different models in Section~\ref{sec:results}
and conclude in Section~\ref{sec:discuss}.

\section{The model}

In this paper we consider linear perturbations to a Friedmann-Lema\^{i}tre-Robertson-Walker 
(FLRW) spacetime and work in the flat gauge, in which the line element takes the form
\be 
ds^2=-(1+2\phi)dt^2+2aB_{,i}dx^idt+a^2(t)\delta_{ij}dx^idx^j\,.
\ee
Here $\phi$ is the lapse function, and $B$ is the scalar shear; we consider flat
spatial slices in agreement with current observations \cite{WMAP7}. 
Our index convention is as follows:
$\{\mu, \nu, \ldots\}$ run over 4 space-time indices, $\{i, j, \ldots\}$ denote the 3 spatial indices,
$\{\alpha, \beta,\ldots\}$ are labels for the `fluids' (including matter and radiation and the kinetic and
potential fluids attributed to the scalar fields -- see later) and $\{A, B, \ldots\}$ label the scalar fields.

In this work we model reheating perturbatively, following the elementary theory 
of reheating developed in Refs.~\cite{Dolgov:1982th, Abbott:1982hn}. This amounts 
to including an additional friction term, $\frac{1}{2}\Gamma \dot{\vp}$, in the equation of motion of the scalar
field during reheating, i.e.\footnote{Note that our decay constants differ from those in, e.g., 
Refs.~\cite{Choi:2008et, Jain:2009ep, Leung:2012ve} 
by a factor of $1/2$. This difference
arises due to the introduction of the decay constants in the fluid conservation equations as 
opposed to the field evolution equations.}
\be 
\ddot{\vp}+(3H+\frac{1}{2}\Gamma)\dot{\vp}+U_{,\vp}=0\,.
\ee
However, this approximation is only valid when the field is rapidly oscillating about its minimum,
and the additional friction term should not be present during inflation.
In our analysis, we ensure that the approximation is valid by only allowing the decay
constants for a particular field to be non-zero once that field has passed through
its potential minimum and is taking part in reheating (see Sec.~\ref{sec:num}).
We describe the system in more detail in the following.

\subsection{Background}

We now study the evolution of the various fluid components, considering
a system with matter and radiation fluids (denoted with a subscript ${\rm m}$ and $\gamma$, respectively) along with multiple scalar fields with the
Lagrangian density
\be 
{\cal L}=\frac{1}{2}\sum_A\dot{\varphi}_A^2+U(\varphi_B)\,.
\ee
We adopt the approach of Ref.~\cite{Malik2005} and
split the scalar fields into kinetic fluids, with energy density and
pressure
$
\rho_A=P_A=\frac{1}{2}\dot{\vp}_A^2\,,
$
and a single potential fluid with
$ 
\rho_U=U=-P_U\,.
$
In the background, energy conservation gives one equation for each fluid, $\alpha$,
\be 
\dot{\rho_{\alpha}}+3H(\rho_\alpha+P_\alpha)=Q_\alpha \,,
\ee
where $Q_\alpha$ is the respective energy transfer function.
So, the radiation and matter fluid conservation equations are 
\begin{align}
\dot{\rho_\gamma}+4H\rho_\gamma &=\frac{1}{2}\sum_A \Gag\dot{\vp}_A^2\,,\\
\dot{\rho_{\rm m}}+3H\rho_{\rm m} &= \frac{1}{2}\sum_A \Gam\dot{\vp}_A^2\,,
\end{align}
where we use the equations of state for the radiation and matter: $P_\gamma = \frac{1}{3}\rho_\gamma$, 
$P_{\rm m}=0$.
The fields each have a Klein-Gordon equation coming from energy-momentum conservation
of the kinetic fluids
\begin{align}
\ddot{\vp}_A+\Big[3H+\frac{1}{2}\Big(\Gam+\Gag\Big)\Big]\dot{\vp}_A+U_{,A}=0\,,
\end{align}
and the evolution equation for the potential fluid is satisfied identically. 

\subsection{Linear perturbations}

Considering now the evolution of linear perturbations, 
the energy-momentum conservation equations 
yield the governing equations for each (perfect) fluid, $\alpha$, \cite{Malik:2002jb, Malik2005},
\begin{align}
\dot{\delta\rho_\alpha}&+3H(\delta\rho_\alpha+\delta P_\alpha)
+\frac{\nabla^2}{a}v_\alpha(\rho_\alpha+P_\alpha)\nn\\
&=Q_\alpha \phi + \delta Q_{\alpha}\,,\\
\label{eq:Valpha}
\dot{\delta q_\alpha}&+3H\delta q_\alpha
+(\rho_\alpha + P_\alpha)\phi+\delta P_\alpha -
Q_\alpha V)=0\,,
\end{align}
where the momentum perturbation, $\delta q_\alpha = a(v_\alpha+B)(\rho_\alpha + P_\alpha)$ 
and $v_\alpha$ is the three-velocity of the $\alpha$ fluid. $V$ is
defined as
\be 
V=\frac{1}{\rho+P}\sum_\alpha \delta q_\alpha\,.
\ee
For our system this becomes
\be 
V=\frac{\delta q_{\rm m} + \delta q_\gamma
+\sum_A\dot{\vp}_A^2 V_A}{\rho_{\rm m}+\frac{4}{3}\rho_\gamma+\sum_A\dot{\vp}_A^2}\,,
\ee
where the evolution of $\delta q_{\rm m}$ and $\delta q_{\gamma}$ is governed by
their respective Eq.~(\ref{eq:Valpha}) and, 
for the fields, we have
$
V_A=-\delta\vp_A / \dot{\vp}_A
$
giving the expression
\be 
V=\frac{\rho_{\rm m}V_{\rm m}+\frac{4}{3}\rho_\gamma V_\gamma
-\sum_A\dot{\vp}_A \delta\vp_A}{\rho_{\rm m}+\frac{4}{3}\rho_\gamma+\sum_A\dot{\vp}_A^2}\,.
\ee
Then, on computing the transfer functions, the fluid energy conservation equations are
\begin{align}
\dot{\delta\rho_\gamma}&+4H\delta\rho_\gamma
+\frac{4}{3}\frac{\nabla^2}{a}v_\gamma\rho_\gamma  
=\sum_A\Gag\Big(\dot{\vp}_A\dot{\delta\vp}_A-\frac{1}{2}\dot{\vp}_A^2\phi\Big)\,,
\end{align}
\begin{align}
\label{eq:mcons}
\dot{\delta\rho_{\rm m}}&+3H\delta\rho_{\rm m}
+\frac{\nabla^2}{a}v_{\rm m}\rho_{\rm m}  
=\sum_A\Gam\Big(\dot{\vp}_A\dot{\delta\vp}_A-\frac{1}{2}\dot{\vp}_A^2\phi\Big)
\,,
\end{align}
and momentum conservation gives
\begin{align}
\label{eq:Vg}
\dot{\delta q_\gamma}+ 3H\delta q_\gamma
+\frac{4}{3}\rho_\gamma\phi
+\frac{\delta\rho_\gamma}{3}
-\frac{V}{2}\sum_A \Gag\dot{\vp}_A^2
=0\,,
\end{align}
\begin{align}
\label{eq:Vm}
\dot{\delta q_{\rm m}}
+3H\delta q_{\rm m}+\rho_{\rm m}\phi 
-\frac{V}{2}\sum_A\Gam\dot{\vp}_A^2=0\,.
\end{align}

In order to obtain the set of equations in a closed form we choose the base variables 
for the fluids to be the momentum and energy density perturbations, $\delta q_\alpha$
and $\delta \rho_\alpha$.
We then need to invoke the perturbed Einstein field equations
which give the two required equations \cite{Malik2005},
\begin{align}
\label{eq:eins00}
3H^2\phi+H\frac{\nabla^2}{a}B=-4\pi G \delta\rho\,,\\
\label{eq:eins0i}
H\phi=-4\pi G (\rho+P)V\,,
\end{align}
where the total energy density perturbation is $\delta\rho=\sum_\alpha \delta\rho_\alpha$.
Considering the matter conservation equation, Eq.~(\ref{eq:mcons}), we can use the fact that
$v_{\rm m}={V_{\rm m}} / {a}-B$, to write
\begin{align}
\dot{\delta\rho_{\rm m}}&+3H\delta\rho_{\rm m}
+\frac{\nabla^2}{a^2}\delta q_{\rm m}
-\frac{\nabla^2}{a}B\rho_{\rm m} \nn\\
&=\sum_A\Gam\Big(\dot{\vp}_A\dot{\delta\vp}_A-\frac{1}{2}\dot{\vp}_A^2\phi\Big)
\,.
\end{align}
Then, from Eq.~(\ref{eq:eins00}), we have 
\be 
\frac{\nabla^2}{a}B=-\frac{4\pi G}{H}\delta\rho-3H\phi\,,
\ee
and so
\begin{align}
\label{eq:drhom}
\dot{\delta\rho_{\rm m}}&+3H\delta\rho_{\rm m}
+\frac{\nabla^2}{a^2}\delta q_{\rm m}+\frac{4\pi G}{H} \rho_{\rm
m}\delta\rho+3H\phi\rho_{\rm m}\nn \\
&=\sum_A\Gam\Big(\dot{\vp}_A\dot{\delta\vp}_A-\frac{1}{2}\dot{\vp}_A^2\phi\Big)
\,,
\end{align}
where the total energy density is \cite{MW2008, Hwang:2001fb}
%
%
%
\be 
\delta\rho=\delta\rho_{\rm m}+\delta\rho_\gamma
+\sum_A\Big(\dot{\vp}_A\dot{\delta\vp}_A-\dot{\vp}_A^2\phi+U_{,A}
\delta\vp_A\Big)\,.
\ee
We obtain a similar equation for the radiation fluid
\begin{align}
\label{eq:drhog}
\dot{\delta\rho_{\gamma}}&+4H\delta\rho_{\gamma}
+\frac{\nabla^2}{a^2}\delta q_\gamma
+\frac{16\pi G}{3H} \rho_{\gamma}\delta\rho+4H\phi\rho_{\gamma}\nn \\
&=\sum_A\Gag\Big(\dot{\vp}_A\dot{\delta\vp}_A-\frac{1}{2}\dot{\vp}_A^2\phi\Big)
\,.
\end{align}
Finally, to close the system we can write the lapse function in terms of fluid and field variables, using Eq.~(\ref{eq:eins0i}),
as
\be 
\label{eq:phi}
\phi=-\frac{4\pi G}{H}\Big(\delta q_{\rm m} + \delta q_\gamma
-\sum_A\dot{\vp}_A\delta\vp_A\Big)\,.
\ee

Evolution equations for the scalar field perturbations can be obtained using the energy conservation equation
with the energy density of each field given by
$
\delta\rho_A=\dot{\vp}_A\dot{\delta\vp}_A-\dot{\vp}_A^2\phi\,.
$
We obtain a modified Klein-Gordon equation for the $\vp_A$ field:
\begin{align}
&\ddot{\delta\vp}_A+\Big[3H+\frac{1}{2}\Big(\Gam+\Gag\Big)\Big]\dot{\delta\vp}_A
+\sum_j U_{,A B}\delta\vp_B\nn\\
&+\Big[2U_{,A}+\frac{1}{2}\dot{\vp}_A\Big(\Gam+\Gag\Big)\Big]\phi 
-\dot{\vp}_A\dot{\phi}
+\frac{\nabla^2}{a}v_A\dot{\vp}_A =0\,.
\end{align}
Then, using the field equations the Klein-Gordon equation for each field $\vp_A$ becomes
\begin{align}
\label{eq:KGA}
&\ddot{\delta\vp}_A+\Big[3H+\frac{1}{2}\Big(\Gam+\Gag\Big)\Big]\dot{\delta\vp}_A
-\frac{\nabla^2}{a^2}\delta\vp_A\nn\\
&+\sum_B U_{,AB} \delta\vp_B + \Big[2U_{,A} +
\frac{1}{2}\dot{\vp}_A\Big(\Gam+\Gag\Big)+3H\dot{\vp_A}\Big] \phi\nn\\
&-\dot{\vp}_A\dot{\phi}+\frac{4\pi G}{H}\dot{\vp}_A\delta\rho =0\,.
\end{align}
where, as above, the lapse function, $\phi$, can be replaced by using Eq.~(\ref{eq:phi}). \\

The closed set of equations to be solved is therefore Eqs.~(\ref{eq:drhog}), (\ref{eq:drhom}), (\ref{eq:Vg}), (\ref{eq:Vm})
for the fluids, and Eqs.~(\ref{eq:KGA}), for the fields, along with the expression for $\phi$, from
Eq.~(\ref{eq:phi}).\\

\subsection{Non-adiabatic perturbations}

For the remainder of the paper, we consider the 
special case of a system containing two scalar fields $\vp$ and $\chi$, though we note that
the expressions can be readily extended to the multiple field case.

In a general fluid system, one can define the non-adiabatic pressure perturbation by expanding the pressure,
as
\be 
\delta P_{\rm nad} \equiv \delta P - \cs \delta\rho\,,
\ee
where the adiabatic sound speed is defined as $\cs=\dot{P}/\dot{\rho}$. The initial non-adiabatic pressure
perturbation in this system containing two scalar fields is given in terms of the field variables by
\cite{Huston:2011fr}
\begin{align}
\label{eq:dPnfields}
\delta P_{\rm nad} & = \frac{8\pi
G}{3H^2}(U_{,\vp}\dot{\vp}+U_{,\chi}\dot{\chi})(\dot{\vp}\delta\vp+\dot{\chi}\delta\chi)\\
&-2(U_{,\vp}\delta\vp+U_{,\chi}\delta\chi)\nn\\
&-\frac{2}{3H}\frac{(U_{,\vp}\dot{\vp}+U_{,\chi}\dot{\chi})}{(\dot{\vp}^2+\dot{\chi}^2)}
\Big[\dot{\vp}\dot{\delta\vp}+\dot{\chi}\dot{\delta\chi}+U_{,\vp}\delta\vp+U_{,\chi}\delta\chi\Big]\,.\nn
\end{align}
This can then be compared to the pressure perturbation which, for a system of two fields, is 
\begin{align}
\delta P=\dot{\vp}\dot{\delta\vp}+\dot{\chi}\dot{\delta\chi}-(\dot{\vp}^2+\dot{\chi}^2)\phi
-U_{,\vp}\delta\vp-U_{,\chi}\delta\chi\,.
\end{align}
%
%
%
%
When the fields have decayed into the fluids, after reheating, the non-adiabatic pressure perturbation is
\be 
\label{eq:dPnfluids}
\delta P_{\rm nad}=\frac{1}{3}\delta\rho_\gamma\Bigg[1-\frac{\dot{\rho}_\gamma}
{\dot{\rho}_{\rm m}+\dot{\rho}_\gamma}\Bigg]
-\frac{1}{3}\frac{\dot{\rho_\gamma}\delta\rho_{\rm m}}{\dot{\rho}_{\rm m}+\dot{\rho}_\gamma}\,,
\ee
where we have used the fact that
\be 
\cs=\sum_\alpha\frac{\dot{\rho}_\alpha}{\dot{\rho}}c_\alpha^2\,.
\ee
This is the non-adiabatic pressure perturbation which we use to calculate the entropy perturbation remaining after
reheating for different potentials in the next section. Throughout the reheating phase, when
the fields are decaying into the fluids, the non-adiabatic pressure perturbation will have
components from both Eqs. (\ref{eq:dPnfields}) and (\ref{eq:dPnfluids}).

\section{Numerical method}
\label{sec:num}

The results in Section~\ref{sec:results} were derived using the {\sc Pyflation} numerical package \cite{pyflation}.
The implementation of the inflationary calculation is described in detail in Refs.~\cite{Huston:2011fr,hustonmalik2}.

The {\sc Pyflation} package has been updated to include the evolution equations of radiation and matter fluids 
at background and first order in perturbations. These fluids initially are set to zero and there is no transfer
from the inflationary fields until each field has reached its minimum value for the first time. 

Reheating is assumed to end when the energy density of the scalar fields is reduced to below some specified
fraction of the total energy density of the system. This value can be set as a parameter
($\rho_{\rm limit}$ in the following) and for the results 
below is either $10^{-5}$ or $10^{-10}$. 
The scalar fields are assumed to play no further role in the evolution after this point and
are set to zero from then onwards. This allows the code to avoid wasting time following extremely small oscillations
of the negligible scalar fields. The perturbations of the scalar fields are also set to zero to maintain the consistency of the
perturbative scheme.
Computations of the non-adiabatic pressure perturbation 
and other derived quantities now include the radiation and matter fluids
as described above.

\section{Results}
\label{sec:results}
In this section we present our results. We focus on the three inflationary potentials we studied in 
Ref.~\cite{Huston:2011fr}, namely the double quadratic, double quartic and product exponential
models. The decay constants are crucial to our result, since they control how quickly the fields
decay into the radiation and matter. We will focus on constraints for these parameters for 
each specific model below. However, one constraint common to all models is that, 
in order for matter to be subdominant
until matter-radiation equality, we require \cite{Gupta:2003jc, Choi:2008et}
\be 
\frac{\Gamma_{\rm m}}{\Gamma_{\gamma}}\lesssim
\Bigg[\frac{\Omega_{\rm m}}{\Omega_{\gamma}}\Bigg]_{\rm decay}
=\frac{T_{\rm eq}}{T_{\rm decay}}<10^{-6}\,.
\ee
In the following we therefore ensure that this bound is satisfied by taking
 $\Gamma^A_{\rm m} = 10^{-6} \Gamma^A_{\gamma}$
for each scalar field, $A$.

\subsection{Double quadratic inflation}
\label{sec:hybquad}

\begin{figure}
 \centering
 \includegraphics[width=0.5\textwidth]{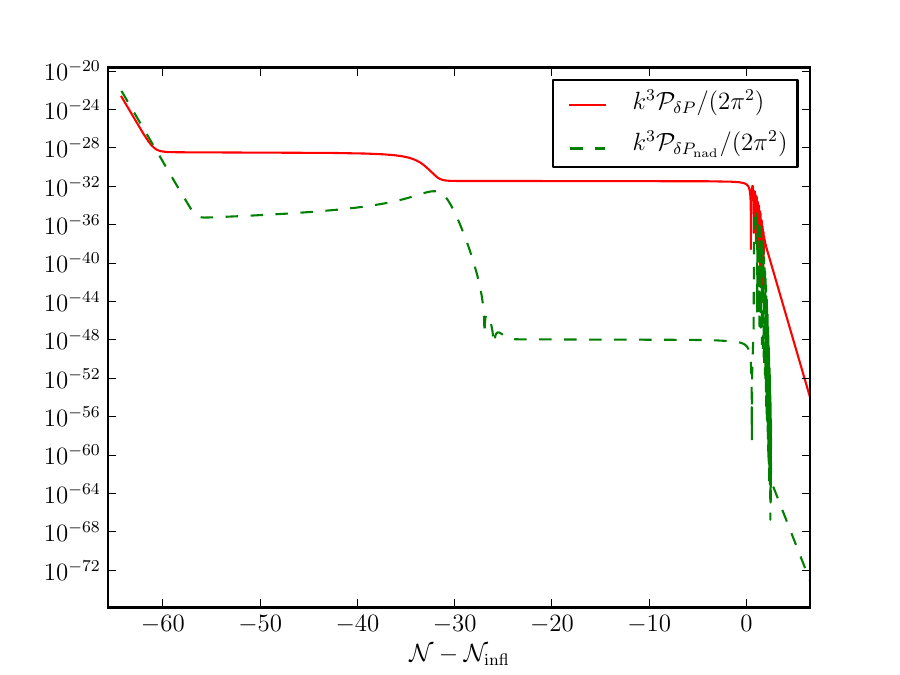}
 \caption{A comparison between the power spectra of $\delta P$ (red solid line) and 
 $\delta P_{\rm nad}$ (green dashed line) for the double quadratic potential at the 
 {\sc Wmap} pivot scale.}
 \label{fig:hybquad_r10_dPnaddP_N_reh-large}
\end{figure}

The first model we consider is the two-field generalisation of quadratic
inflation, the double quadratic inflationary model, with the potential
\be 
V(\vp,\chi)=\frac{1}{2}m_\vp^2\vp^2+\frac{1}{2}m_\chi^2\chi^2\,.
\ee
We use the parameter values from Ref.~\cite{Huston:2011fr}, choosing $m_\chi=7m_\vp$
and setting $m_\vp=1.395\times10^{-6}M_{\rm PL}$. In Fig.~\ref{fig:hybquad_r10_dPnaddP_N_reh-large}
we plot the power spectra of the pressure perturbation, $\delta P$ (red solid line) and the
non-adiabatic pressure perturbation, $\delta P_{\rm nad}$ (green dashed line). In all graphs in this section
results are plotted against number of efoldings after the end of inflation (i.e. inflation ends
at the origin on the $x$-axis). We can see that, in this model, as inflation ends the spectrum of the
non-adiabatic pressure perturbation is many order of magnitude smaller than that of the
pressure perturbation, as presented in Ref.~\cite{Huston:2011fr}. 

\begin{figure}
 \centering
 \includegraphics[width=0.5\textwidth]{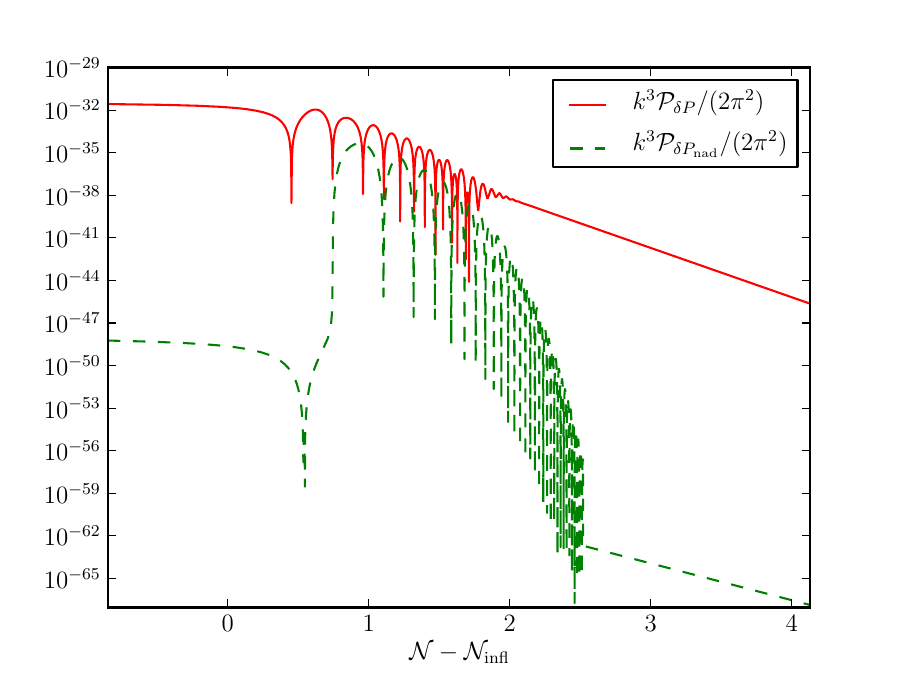}
 \caption{A comparison between the power spectra of $\delta P$ (red solid line) and 
 $\delta P_{\rm nad}$ (green dashed line) for the double quadratic potential during the reheating
 phase only.}
 \label{fig:hybquad_r10_dPnaddP_N_reh_zoom-large}
\end{figure}

\begin{figure}
 \centering
 \includegraphics[width=0.5\textwidth]{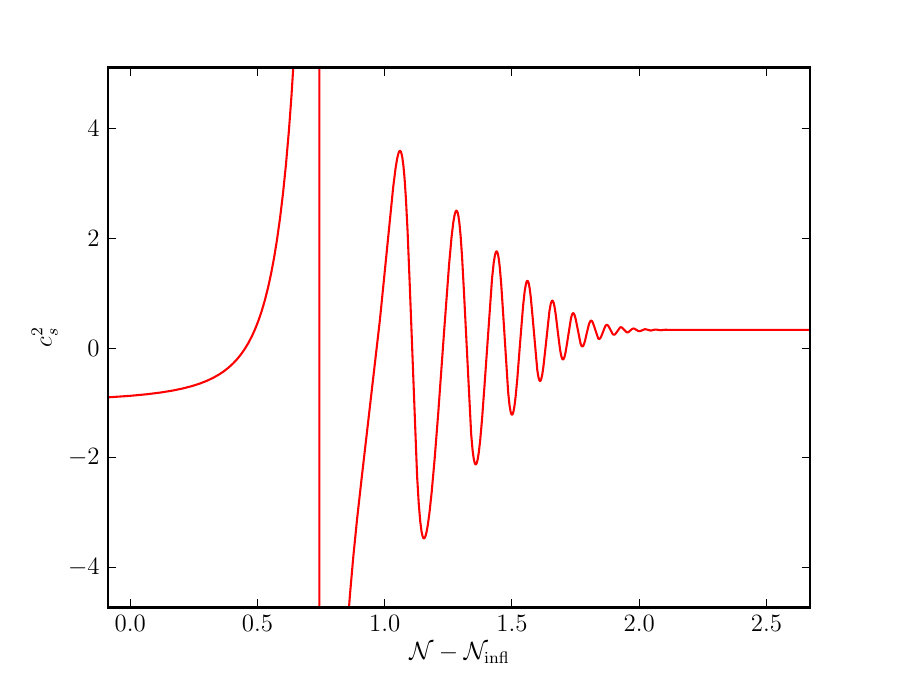}
 \caption{The adiabatic sound speed of the system for the double quadratic inflation model.}
 \label{fig:hybquad_csq_N_reh-large}
\end{figure}

In Fig.~\ref{fig:hybquad_r10_dPnaddP_N_reh_zoom-large} we zoom in to the last few efolds of the 
previous plot, and consider only the evolution through reheating. We can see that during the initial phases
of reheating, the isocurvature\footnote{Note that, in this paper, we will
use the terms `isocurvature' and `non-adiabatic pressure' 
interchangeably.} perturbation sharply increases, due to the generation of radiation and matter
fluids and the corresponding fluctuation of the adiabatic sound speed of the total fluid 
(as seen in Fig.~\ref{fig:hybquad_csq_N_reh-large}).

However, the non-adiabatic pressure perturbations decay faster than the pressure perturbation during 
the period of reheating. When the fields are turned off and reheating is over, approximately 2.5 efolds 
after the end of inflation, the non-adiabatic pressure perturbation is around twenty orders of magnitude 
smaller than the pressure perturbation. Thus, we conclude that the reheating phase
does not change the prediction that the isocurvature is negligible in the double quadratic inflationary model.

\begin{figure}
 \centering
 \includegraphics[width=0.5\textwidth]{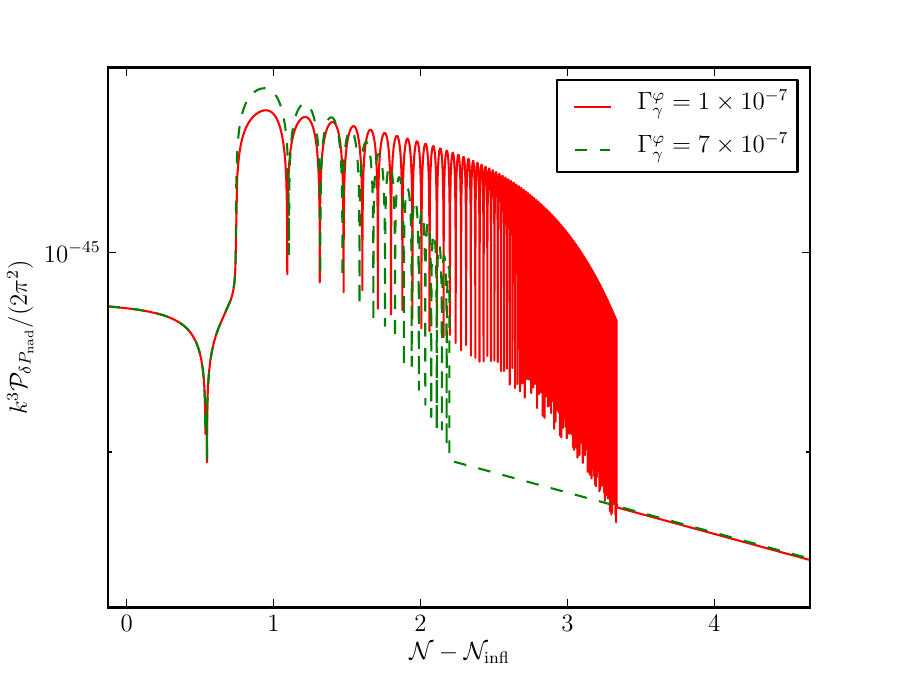}
 \caption{The $\delta P_{\rm nad}$ power spectrum for different values of the decay constant.}
 \label{fig:hybquad_compareGammas_dPnad_N_reh-large}
\end{figure}

In order to model reheating for this potential we have had to specify the decay constants. 
The results in Figs. \ref{fig:hybquad_r10_dPnaddP_N_reh-large} and \ref{fig:hybquad_r10_dPnaddP_N_reh_zoom-large}
are obtained by choosing the decay constants $\Gamma^\vp_\gamma$ and $\Gamma^\chi_\gamma$
to be equal and set to $7 \times 10^{-7}M_{\rm PL}$. Allowing these to be different from one another and setting
$\Gamma_\gamma^\vp=1\times 10^{-7}M_{\rm PL}$ while keeping 
$\Gamma_\gamma^\chi=7\times 10^{-7}M_{\rm PL}$,
we can see from Fig.~\ref{fig:hybquad_compareGammas_dPnad_N_reh-large} that this simply 
means that reheating takes longer, while the spectrum obtained at the end is the same magnitude.


\begin{figure}
 \centering
 \includegraphics[width=0.5\textwidth]{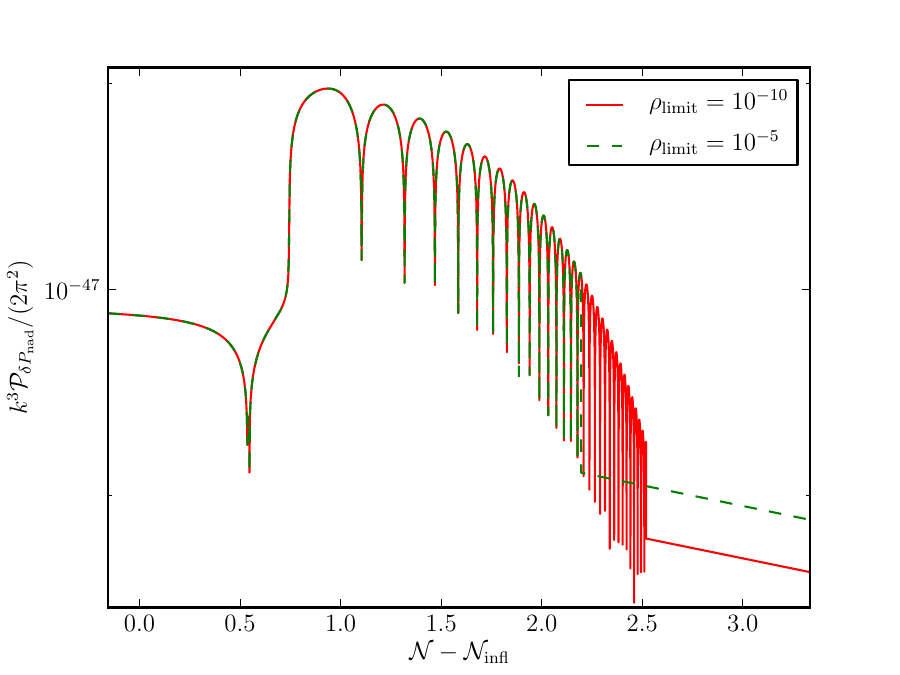}
 \caption{A comparison of the $\delta P_{\rm nad}$ power spectrum for two values of 
 $\rho_{\rm limit}$, the value of the field energy density below which the field is turned off.}
 \label{fig:hybquad_compare_dPnad_N_reh-large}
\end{figure}

Furthermore, as described in Section~\ref{sec:num}, in the numerical procedure we drop a 
field from the dynamics when its energy density becomes smaller than a certain fraction
of the total energy density. In Fig.~\ref{fig:hybquad_compare_dPnad_N_reh-large} we plot
 the non-adiabatic pressure power spectrum for two values of this ratio, $\rho_{\rm limit}=10^{-5}, 10^{-10}$, and we can see that allowing the fields
 to take part in the simulation for longer, i.e. choosing a smaller value for
 $\rho_{\rm limit}$, results in a non-adiabatic pressure perturbation
 spectrum with a smaller amplitude.

\begin{figure}
 \centering
 \includegraphics[width=0.5\textwidth]{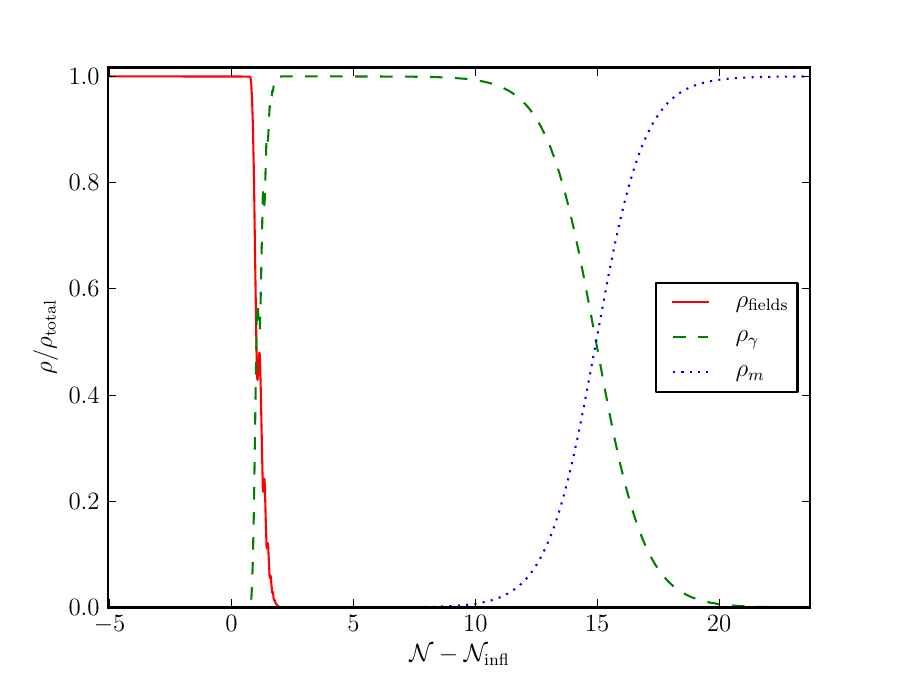}
 \caption{The evolution of the energy density of the fields (red solid line), 
 radiation (green dashed line) and matter (blue dotted line) as a fraction 
 of the total energy density.}
 \label{fig:hybquad_rhofrac_N_reh-large}
\end{figure}

Finally, we can confirm that the Universe reheats properly in this model by considering the 
complete background evolution. Plotted in Fig.~\ref{fig:hybquad_rhofrac_N_reh-large} are the 
fractional energy densities of the constituent components of the cosmic fluid compared
to the total energy density. We can see that as the model reheats, the proportion of
energy density in the fields drops to zero and is converted into radiation and a small amount of
matter. At this point in time, the adiabatic sound speed of the system is $c_{\rm s}=1/3$, 
as shown in Fig.~\ref{fig:hybquad_csq_N_reh-large}, as the system enters a radiation dominated era, followed by 
a matter dominated era towards the present day. We do not consider the 
evolution of the perturbations for the entire history of the Universe since our approximations
are not valid, and we do not include dark energy perturbations. 
However, obtaining the correct background evolution into the matter
dominated era is a good confirmation that the model is sufficiently accurate.

\subsection{Double quartic inflation}

The next model we consider is a special case of hybrid inflation, the double quartic model
whose potential takes the form 
\be 
V(\vp,\chi)=\Lambda^4\Bigg[
\Bigg(1-\frac{\chi^2}{v^2}\Bigg)^2+\frac{\vp^2}{\mu^2}+\frac{2\vp^2\chi^2}{\vp_{\rm c}^2v^2}\Bigg]\,,
\ee
where $v, \mu$ and $\vp_c$ are constants.
In order to compare with our previous results, we use the same parameter values as in Ref.~\cite{Huston:2011fr}.

\begin{figure}
 \centering
 \includegraphics[width=0.5\textwidth]{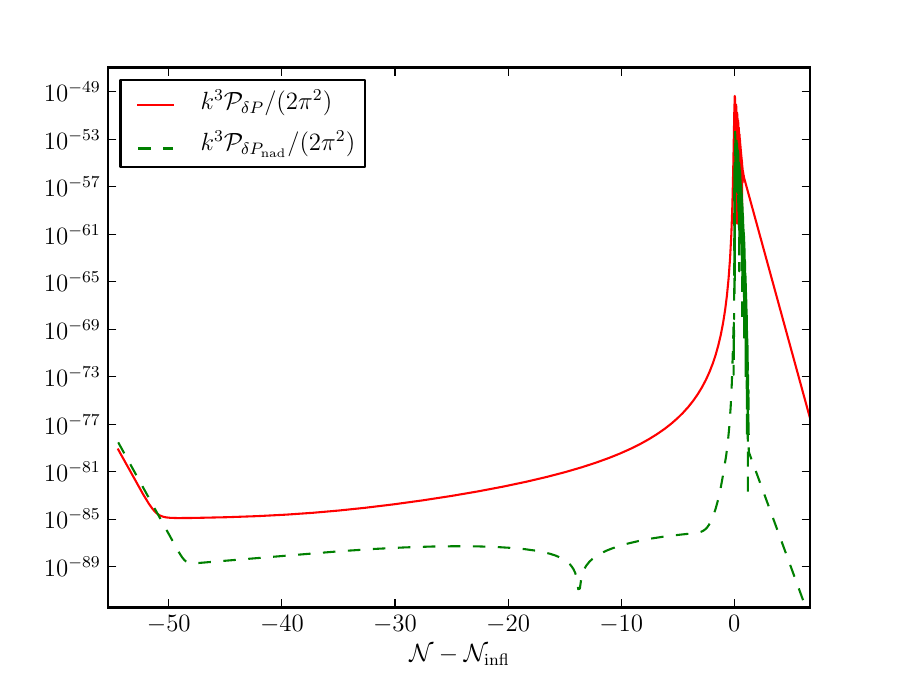}
 \caption{A comparison between the power spectra of $\delta P$ (red solid line) and 
 $\delta P_{\rm nad}$ (green dashed line) for the double quartic inflationary potential at the 
 {\sc Wmap} pivot scale.}
 \label{fig:hybquart_r10_dPnaddP_N_reh-large}
\end{figure}

\begin{figure}
 \centering
 \includegraphics[width=0.5\textwidth]{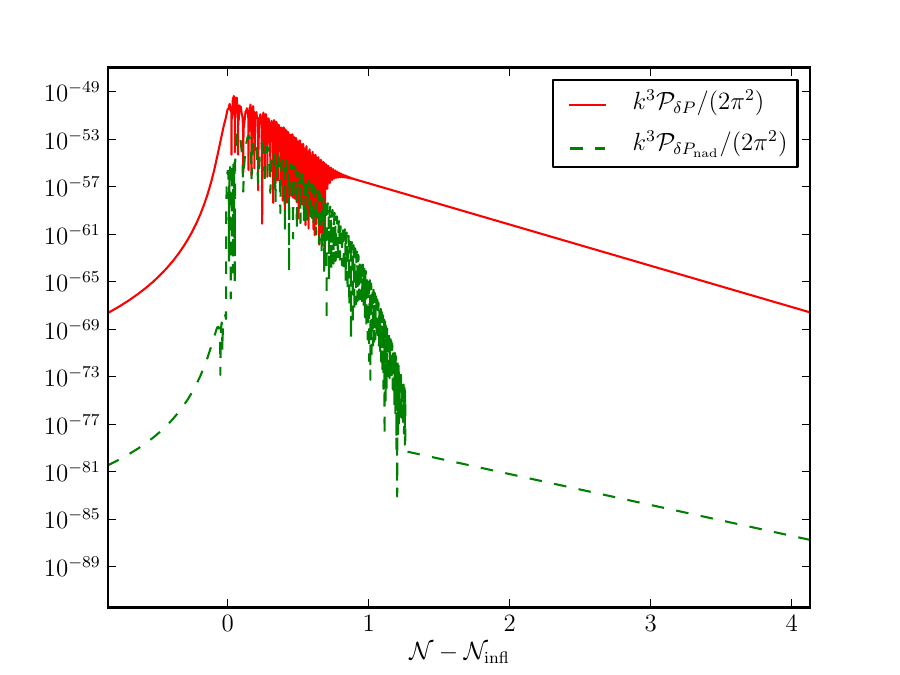}
 \caption{A comparison between the power spectra of $\delta P$ (red solid line) and 
 $\delta P_{\rm nad}$ (green dashed line) zoomed in to the reheating phase of the 
 double quartic inflationary potential.}
 \label{fig:hybquart_r10_dPnaddP_N_reh_zoom-large}
\end{figure}

Plotted in Fig.~\ref{fig:hybquart_r10_dPnaddP_N_reh-large} is the comparison
between the power spectra of the pressure and non-adiabatic pressure perturbations
as a function of time. The difference from the double quadratic potential during the inflationary epoch 
is apparent, and we can see the sharp increase of the isocurvature towards the end of inflation. 
When focussing on the reheating phase, as presented in Fig.~\ref{fig:hybquart_r10_dPnaddP_N_reh_zoom-large},
we can see that, while the non-adiabatic pressure spectrum is within a few orders of magnitude
of that of the pressure perturbation at the end of inflation, as reheating progresses, the
isocurvature decays more rapidly. When this model has fully reheated, we can see that the 
isocurvature spectrum is again negligible, roughly twenty orders of magnitude smaller
than the spectrum of the pressure perturbation.

The decay parameters used for this model follow Ref.~\cite{Choi:2008et} closely, and are
$\Gamma^\vp_\gamma=\Gamma^\chi_\gamma=3\times10^{-11}M_{\rm PL}$ and 
$\Gamma^\vp_{\rm m}=\Gamma^\chi_{\rm m}=3\times10^{-17}M_{\rm PL}$. We have checked that, as
in the previous case, varying these parameter values does not drastically change the results.

In a similar way to in the previous model, we can show that the reheating of the double quartic model
also produces the correct background evolution of the universe, i.e. the fields reheat
resulting in a radiation dominated epoch followed by matter domination. 

%
%
%
%
%
%
%
%

\subsection{Product exponential}

\begin{figure}
 \centering
 \includegraphics[width=0.5\textwidth]{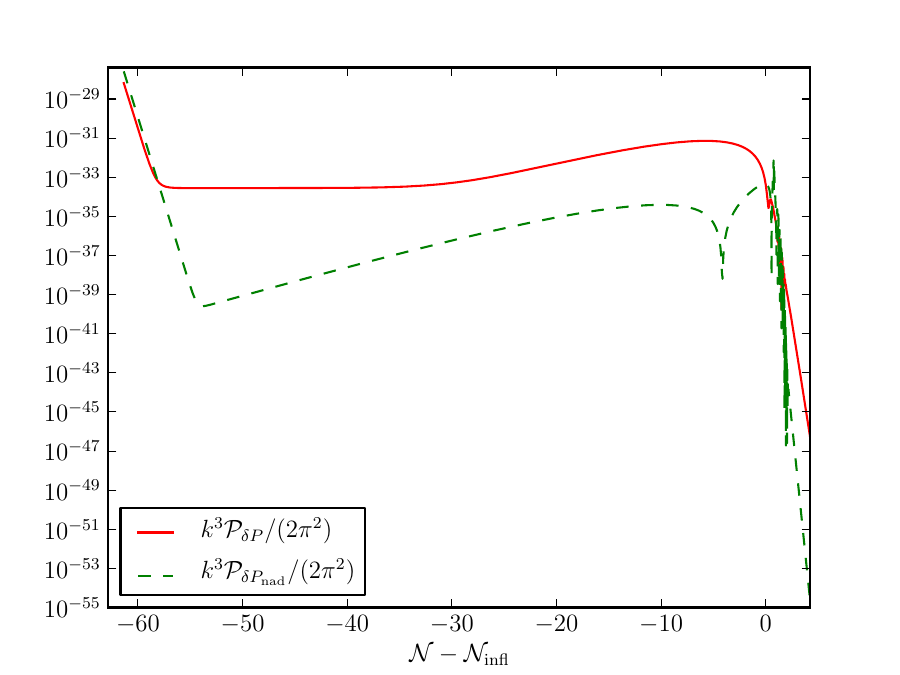}
 \caption{The power spectra of $\delta P$ (red solid line) and $\delta P_{\rm nad}$ (green dashed line)
 for the product exponential potential at the {\sc Wmap} pivot wavenumber.}
 \label{fig:prodexp_r10_dPnaddP_N_reh-large}
\end{figure}

Finally, we study the product exponential potential, a model popular in the literature \cite{Choi:2008et, Elliston:2011dr}
due to its tendency to source large non-gaussianities and which has a large isocurvature perturbation present at
the end of inflation. The potential is
\begin{equation}
 V = V_0 \vp^2 e^{-\lambda \chi^2}\,,
\end{equation}
%
%
and we use parameter values as in Ref.~\cite{Huston:2011fr}. In Fig.~\ref{fig:prodexp_r10_dPnaddP_N_reh-large}
we show the results for this model, plotting the power spectrum of the pressure perturbation
compared with the power spectrum of the non-adiabatic pressure perturbation. 
In this figure we can see the characteristic growth of 
the non-adiabatic pressure perturbation power spectrum, peaking
around the end of inflation where it is then larger than that of the pressure perturbation as inflation ends.
This was shown in Ref.~\cite{Huston:2011fr}. Choosing the decay constants used in Ref.~\cite{Leung:2012ve},
we investigate the evolution of this isocurvature through reheating.

%

\begin{figure}
 \centering
 \includegraphics[width=0.5\textwidth]{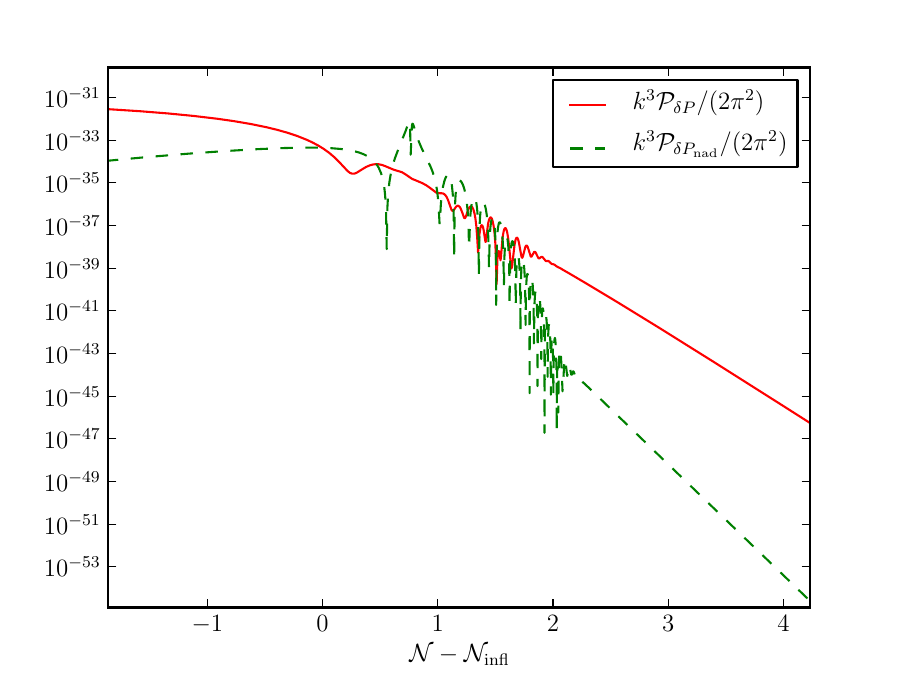}
 \caption{A comparison between the power spectra during reheating of $\delta P$ (red solid line) and $\delta P_{\rm nad}$
  (green dashed line) for the product exponential potential.}
 \label{fig:prodexp_r10_dPnaddP_N_reh_zoom-large}
\end{figure}

In Fig.~\ref{fig:prodexp_r10_dPnaddP_N_reh_zoom-large} we compare the pressure and non-adiabatic pressure
perturbation for this model, focussing on the period of reheating. We see that after inflation
ends both power spectra decay. The isocurvature spectrum is larger than that of the pressure perturbation
immediately after inflation, but decays faster. When reheating ends the spectra stop oscillating and 
enter a period of monotonic decay. 
The non-adiabatic pressure spectrum is around five orders of magnitude 
smaller than the spectrum of pressure perturbations at this time, and continues to decay faster than
the pressure perturbation spectrum.

\section{Discussion}
\label{sec:discuss}

In this paper we have studied the period of reheating, where the scalar fields driving inflation
decay into the standard matter and radiation fluids in the universe today, for three different
two-field inflationary models. Each model contained a canonical kinetic term along with a 
specific form for the potential. We modelled the reheating phase perturbatively, introducing 
constants which allowed for the decay of the two scalar fields into both matter and radiation,
and solved the system fully numerically. Doing so enabled us to describe the evolution of
the non-adiabatic pressure perturbations beyond the inflationary phase, and through 
reheating, thereby extending work previously undertaken in Ref.~\cite{Huston:2011fr}.

We found that the isocurvature perturbations are subdominant in all cases. In the double
quadratic model, the isocurvature is negligible after inflation, and after the reheating 
period remains negligible. Therefore, the prediction of this model is robust, namely that
there is no isocurvature for a double quadratic potential. The double quartic potential 
is a model in which the non-adiabatic pressure spectrum is within a few orders of 
magnitude of the pressure perturbation spectrum when inflation ends. During 
reheating, the isocurvature rapidly decays, and when the fields have decayed into fluids,
the isocurvature spectrum is negligible, around twenty orders of magnitude smaller. 
The product exponential potential is the only model we studied in which the 
non-adiabatic pressure power spectrum is larger than that of the pressure perturbation
for a short time when inflation ends. As the universe reheats, the isocurvature again
decays faster than the pressure. However, when the reheating phase is over, the isocurvature spectrum
is within four orders of magnitude of the pressure perturbation spectrum. 

The product
exponential model is the only potential studied here in which the isocurvature fraction after
reheating is non-negligible. This differs from the prediction directly from inflation,
however shows that isocurvature perturbations remain important in this case.

An important requirement of the model of reheating is that it reproduces the correct
expansion history of the universe. Thus, we ensured that this was true for the models
studied in this paper (see, e.g., Fig.~\ref{fig:hybquad_rhofrac_N_reh-large}). The fact that the isocurvature
is subdominant likely arises from the constraint on the decay parameters ensuring that the 
radiation/matter fraction is correct in order for big bang nucleosynthesis to take place. 
While our model gives us the correct energy density fractions for the history of the universe,
we do not trust its predictions too far into the radiation dominated era, since in order to consider
the period around recombination, we need to use a CMB Boltzmann code. However, this is a good check of the model.
Of course, non-adiabatic pressure perturbations arising from the inflationary/reheating mechanism are
not the only way in which isocurvature can present itself in the universe. At a later time, 
around matter-radiation equality, this relative entropy
between matter and radiation sources an isocurvature perturbation, as studied in Ref.~\cite{Brown:2011dn}.

Throughout this work, 
we made the assumption of constant decay parameters.
However, more general expressions can be obtained, such as through 
the inclusion of a thermal background, leading to time-dependent
decay constants. This is beyond the scope of the current paper, but could
be investigated in future work.

In summary, in this paper we have addressed the problem of how to obtain realistic predictions
for inflationary models which exhibit a large amount of isocurvature when inflation ends,
presenting results for the non-adiabatic pressure perturbation that exists when the universe 
has reheated. Such isocurvature could have important observational consequences, such as
through the generation of vorticity at second order in perturbation theory (see discussion in Ref.~\cite{Huston:2011fr}
and Refs.~\cite{vorticity, Christopherson:2010ek, Christopherson:2010dw}). This will be explored
in a future publication.

\section*{Acknowledgements}
The authors are grateful to Anne Green, Godfrey Leung, 
Karim Malik, David Mulryne, David Seery and Ewan Tarrant for useful discussions. 
AJC thanks the Astronomy Unit at QMUL for hospitality.
IH is supported by the STFC under Grant
ST/G002150/1, and AJC is funded by the Sir Norman Lockyer
Fellowship of the Royal Astronomical Society. 

\bibliography{reheatpapers}

\end{document}